# Synthesis of low-moment CrVTiAl: a potential room temperature spin filter

G. M. Stephen[1], I. McDonald[2,3], B. Lejeune[2], L. H. Lewis[2,3], and D. Heiman[1]

[1]*Department of Physics, Northeastern University, Boston, Massachusetts, 02115, USA*

[2]*Department of Chemical Engineering, Northeastern University, Boston, Massachusetts, 02115, USA*

[3]*Department of Mechanical and Industrial Engineering, Northeastern University, Boston, Massachusetts, 02115, USA*

Abstract

The efficient production of spin-polarized currents at room temperature is fundamental to the advancement of spintronics. Spin-filter materials – semiconductors with unequal band gaps for each spin channel – can generate spin-polarized current without the need for spin-polarized contacts. In addition, a spin-filter material with zero magnetic moment would have the advantage of not producing strong fringing fields that would interfere with neighboring electronic components and limit the volume density of devices. The quaternary Heusler compound CrVTiAl has been predicted to be a zero-moment spin-filter material with a Curie temperature in excess of 1000 K. In this work, CrVTiAl has been synthesized with a lattice constant of $a = 6.15$ Å. Magnetization measurements reveal an exceptionally low moment of $\mu = 2.3 \times 10^{-3}$ $\mu_B$/f.u. at a field of $\mu_0 H = 2$ T, that is independent of temperature between $T = 10$ K and 400 K, consistent with the predicted zero-moment ferrimagnetism. Transport measurements reveal a combination of metallic and semiconducting components to the resistivity. Combining a zero-moment spin-filter material with nonmagnetic electrodes would lead to an essentially nonmagnetic spin injector. These results suggest that CrVTiAl is a promising candidate for further research in the field of spintronics.

Future spintronic devices rely on the production of spin-polarized currents at room temperature.[1–4] Spin-polarized currents can be generated several ways: by passing an unpolarized current through a ferromagnetic contact, using the spin-Hall or spin-Seebeck effects, by ballistic and hot electron injection, by employing spin-polarized materials such as half-metals, or using a *spin-filter* material. Half metals[5–9] and spin-filter (SF) materials[10,11] are especially suitable for spin injectors that are based on magnetic tunnel junctions (MTJs).[12] Furthermore, a spin injector using a spin-filter material is simpler, as it does not require magnetic electrodes.

Spin-filter materials are magnetic semiconductors where the two spin states have unequal band gaps, shown in Fig. 1. When unpolarized electrons tunnel through a thin SF barrier they encounter a potential barrier that is lower for one spin state. As the tunneling probability increases exponentially with decreasing barrier height, the current for one spin direction will be much larger than for the other spin direction, thus creating a spin-polarized current. In practice, a spin filter is a simple MTJ-based device with the tunneling insulator replaced by the SF material, but more importantly, the magnetic MTJ electrodes are replaced by a nonmagnetic metal.[13] This polarizing effect can be varied by applying a voltage to the spin-filter electrodes, hence creating a tunable spin valve. Spin filter devices have been fabricated with the magnetic semiconductor EuS[11,14]; unfortunately, its low Curie temperature ($T_C = 16$ K) limits its applicability.

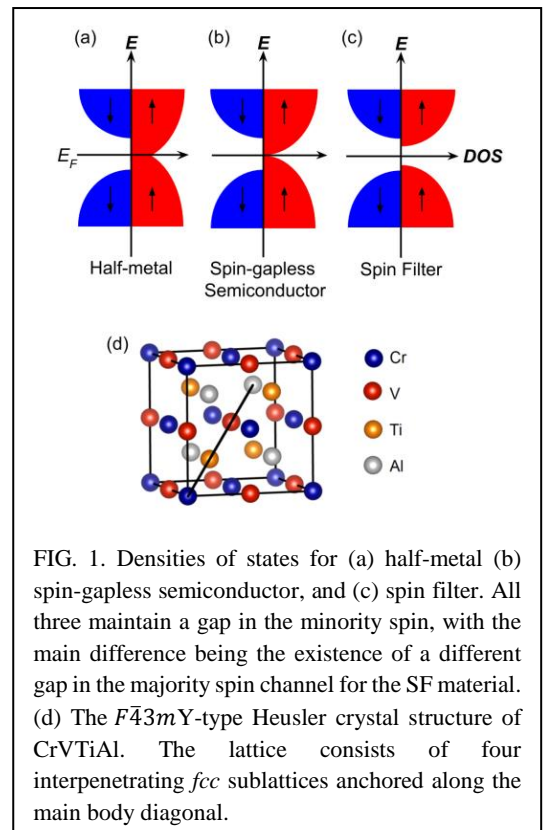

FIG. 1. Densities of states for (a) half-metal (b) spin-gapless semiconductor, and (c) spin filter. All three maintain a gap in the minority spin, with the main difference being the existence of a different gap in the majority spin channel for the SF material. (d) The $F\bar{4}3m$ Y-type Heusler crystal structure of CrVTiAl. The lattice consists of four interpenetrating *fcc* sublattices anchored along the main body diagonal.

Present research on spin injectors has focused primarily on ferromagnetic half metals, which have sizeable moments and strong fringing fields. Half-metallic Heusler compounds such as $Co_2FeSi$[15,16] and $Co_2MnAl$[17] are candidates for spin injection as their electronic structures could provide up to 100 % spin-polarization at the Fermi level. However, the magnetism that is necessary to split the spin-degeneracy in these materials also produces strong fringing fields. A traditional antiferromagnet (AF), while not producing problematic external fields, cannot have a spin-polarized band-structure due to the symmetry in the magnetic moments in the crystal lattice.[18] This dilemma can be remedied with fully-compensated ferrimagnetic materials (FCFiMs) that possess three or more different magnetic sublattices giving rise to a fully-compensated zero total moment. These FCFiMs have the same net-zero moment as an AF, but the asymmetry in the magnetic moments allows spin-polarized behavior. A few half-metallic FCFiMs have been synthesized, including $Mn_2Ru_xGa$.[19,20]

Recently, a number of quaternary Heusler compounds have been predicted by Galanakis *et al.* to have spin-filter properties and Curie temperatures as high as $T_C$ = 2000 K.[21–23] Some of these LiMgPdSn-type Y-structure Heusler materials are also fully compensated with zero net moment – thus combining the magnetic advantages of a FCFiM with the spin-filter function, yet having a Curie temperature well above room temperature. One promising SF material is CrVTiAl, which is predicted to crystallize in a Y-type Heusler structure (space group $F\bar{4}3m$), as shown in Fig. 1 (d). CrVTiAl in this cubic structure is predicted to have a 0.32 eV direct band gap for the spin-up channel and a 0.64 eV indirect gap for the spin-down channel, and is thus a compelling material for investigation.[22]

Polycrystalline bulk samples of CrVTiAl were arc-melted in an Ar environment from stoichiometric quantities of pure (99.99%) elements. Samples were annealed for 1 week at 1000° C in flowing Ar and quenched in water to retain the cubic Heusler structure. Energy dispersive X-ray spectroscopy (EDS) confirms the stoichiometry is within 2 at% of ideal. Structural properties were investigated in an X-ray diffractometer using Cu-Kα radiation. Magnetic and transport measurements were performed using a Quantum Design MPMS SQUID magnetometer at fields up to $\mu_0H$ = 5 T and temperatures from T = 10 to 400 K. Samples for electrical transport measurements were prepared by applying indium contacts to polished slices of the bulk ingot. A van der Pauw sample geometry was fabricated and attached to an electrical probe24 modified for the magnetometer.

The measured and predicted X-ray diffraction (XRD) patterns for a polycrystalline CrVTiAl sample are compared in Fig. 2. The integrated intensities of the measured (200), (220), (400), (422) and (440) Bragg peaks closely match the expected intensities, shown by the red dots, for the cubic Heusler structure within 10%. A lattice constant of $a$ = 6.149 ± 0.004 Å was found, which is close to the predicted lattice constant of 6.20 Å.[22,23] Figure 2 (inset) shows the (111) and (200) peaks, indicating partial L2$_1$ or Y-type order is present in the system, however, the intensities are too weak for reliable quantitative measurements.

Magnetization data shown in Fig. 3 demonstrate a temperature-independent near-zero moment, $\mu$ = 2.3 × 10$^{-3}$ $\mu_B$/f.u. at $\mu_0H$ = 2 T, consistent with the expected generalized Slater-Pauling behavior.[25,26] The temperature and field dependence of the magnetization is similar to that obtained for AF $V_3Al$ by Jamer *et al.*[18]. The linear field dependence of the magnetic moment as well as the temperature-independence of the moment are consistent with previous results reported for polycrystalline AFs.[27] The vanishing moment suggests that the disorder does not affect the zero-moment prediction of the Slater-Pauling model.

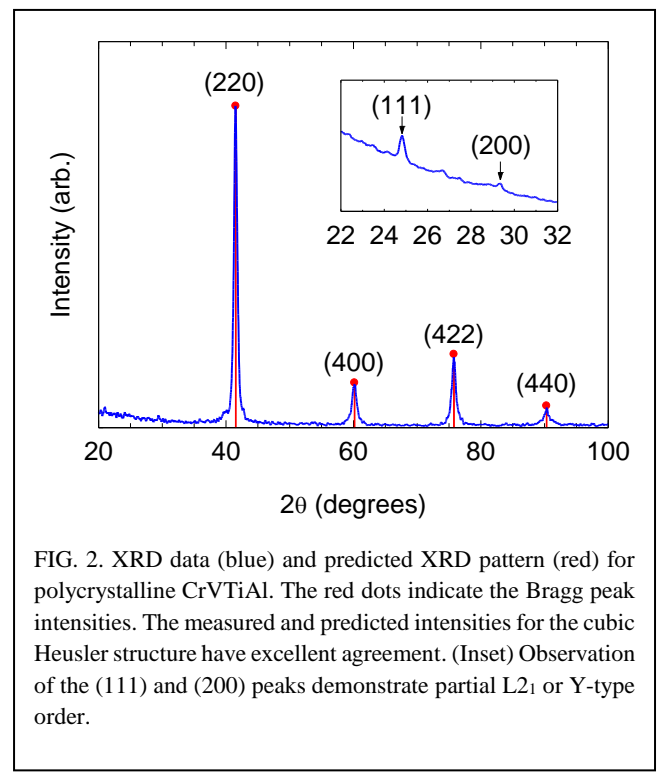

FIG. 2. XRD data (blue) and predicted XRD pattern (red) for polycrystalline CrVTiAl. The red dots indicate the Bragg peak intensities. The measured and predicted intensities for the cubic Heusler structure have excellent agreement. (Inset) Observation of the (111) and (200) peaks demonstrate partial L2$_1$ or Y-type order.

Results of the resistivity measurements are shown in Fig. 4. The sample has a room temperature resistivity of ρ = 210 μΩcm, and is seen to increase with increasing temperature. The resistivity shows metallic-like behavior that is linear in temperature up to T ~ 200 K, above which it diverges from the expected linear behavior. This sublinear behavior can be attributed to an increase in the number of carriers for increasing temperature. Similar behavior is seen in $Mn_2CoAl$ films.[28] The resistivity data may be understood by considering two independent conduction channels, one metallic and the other semiconducting, such that the total conductivity is

$$\sigma(T) = \frac{1}{\rho(T)} = n_m e \mu_m + n_s e \mu_s, \quad (1)$$

where $m$ and $s$ refer to the metallic and semiconducting contributions, respectively. The metallic carrier concentration $n_m$ is taken to be a constant. The inverse mobilities are additive, as they represent series resistances, and are given by

$$\mu_i^{-1} = \alpha_i T + \mu_{0,i}^{-1}, \quad (2)$$

where each conducting channel $i$ will have different values for α and $\mu_{0,i}$. The α term results from electron-phonon scattering while $\mu_{0,i}^{-1}$ corresponds to the mobility due to defects at T = 0 K. The semiconducting carrier concentration ($n_s$) is dominated by a thermal activation energy ΔE, thus

$$n_s = n_0 e^{-\Delta E / k_B T}. \quad (3)$$

Fitting the data in Fig. 4 to this model gives an activation energy of ΔE = 0.16 ± 0.03 eV, and is shown by the solid red curve in Fig. 4. This activation energy is about one-half of the predicted majority band gap of 0.32 eV [23]. This simple model provides the following picture for the resistivity in CrVTiAl: for T < 200 K the carrier density is constant and electron-phonon scattering dominates the resistivity that linearly increases with increasing temperature. For T > 200 K, the resistivity decreases as carriers become thermally activated and are available as current. This thermal activation character is a signature of semiconductor-like energy levels expected for this SF material.

The thermoelectric effect shows the sample to be predominately $p$-type at room temperature, an observation that is consistent with the high DOS in the valence band. The Fermi energy of a $p$-doped sample will lie below the top of the valence band. At low temperatures the system is metallic, so electrons can conduct through empty states in the valence band above the Fermi energy. Increased temperature excites carriers to conduction-band-like states, giving rise to the semiconducting behavior near room temperature.

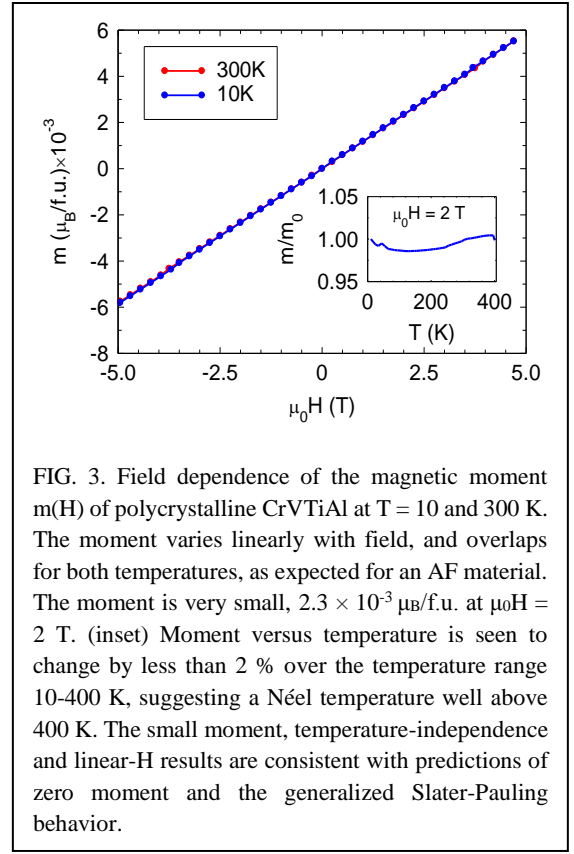

FIG. 3. Field dependence of the magnetic moment m(H) of polycrystalline CrVTiAl at T = 10 and 300 K. The moment varies linearly with field, and overlaps for both temperatures, as expected for an AF material. The moment is very small, 2.3 × 10⁻³ μB/f.u. at μ₀H = 2 T. (inset) Moment versus temperature is seen to change by less than 2 % over the temperature range 10-400 K, suggesting a Néel temperature well above 400 K. The small moment, temperature-independence and linear-H results are consistent with predictions of zero moment and the generalized Slater-Pauling behavior.

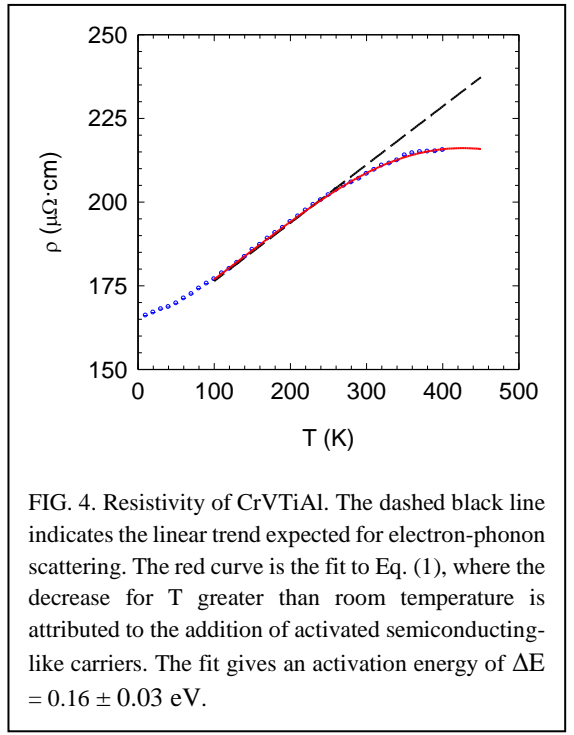

FIG. 4. Resistivity of CrVTiAl. The dashed black line indicates the linear trend expected for electron-phonon scattering. The red curve is the fit to Eq. (1), where the decrease for T greater than room temperature is attributed to the addition of activated semiconducting-like carriers. The fit gives an activation energy of ΔE = 0.16 ± 0.03 eV.

In summary, the quaternary Heusler compound CrVTiAl was synthesized in a partially ordered Y-type or $L2_1$ Heusler structure. Magnetic measurements reveal an exceptionally small temperature-independent magnetic moment of μ ~ $10^{-3}$ μB/f.u., consistent with fully-compensated ferrimagnetism. Electronic transport measurements reveal the presence of thermally-activated carriers consistent with semiconductor-like energy levels. These properties suggest that CrVTiAl is a good candidate for further spintronics research, especially as a room temperature spin filter. Incorporating such a material into a spin injector would have an advantage over traditional spin injectors as the entire device is effectively nonmagnetic.

Acknowledgements

We thank T. Hussey for assistance with magnetometry and M. E. Jamer for helpful conversations. This work was supported by the National Science Foundation grant ECCS-1402738 and by the US Army under grant W911NF-10-2-0098, subaward 15-215456-03-00.